\documentclass[prx,twocolumn,balancelastpage,letterpaper,superscriptaddress]{revtex4-1}
\usepackage{times}

\usepackage{dsfont}
\usepackage{amsmath}
\usepackage{amssymb}
\usepackage{graphicx}
\usepackage{bm}
\usepackage{bbm}
\usepackage{color}
\usepackage{verbatim}
\usepackage{tikz}
\usepackage{hyperref}
\usepackage{float}

\begin{document}

\title{Thermodynamic signatures of edge states in topological insulators}

\author{A. Quelle}
\affiliation{Institute for Theoretical Physics, Center for Extreme Matter and Emergent Phenomena, Utrecht University, Leuvenlaan 4, 3584 CE Utrecht, The Netherlands}
\author{E. Cobanera}
\affiliation{Institute for Theoretical Physics, Center for Extreme Matter and Emergent Phenomena, Utrecht University, Leuvenlaan 4, 3584 CE Utrecht, The Netherlands}
\author{C. \surname{Morais Smith}}
\affiliation{Institute for Theoretical Physics, Center for Extreme Matter and Emergent Phenomena, Utrecht University, Leuvenlaan 4, 3584 CE Utrecht, The Netherlands}

\date{\today}

\begin{abstract}
Topological insulators are states of matter distinguished 
by the presence of symmetry protected metallic boundary states. These edge modes 
have been characterised in terms of transport and spectroscopic measurements, 
but a thermodynamic description has been lacking. The challenge arises because 
in conventional thermodynamics the potentials are required to scale linearly with extensive variables like volume, which does not allow for a general treatment of 
boundary effects. In this paper, we overcome this challenge with Hill thermodynamics. In this extension of 
the thermodynamic formalism, the grand potential is split into an extensive, conventional contribution, and the subdivision potential, 
which is the central construct of Hill's theory. For topologically non-trivial 
electronic matter, the subdivision potential captures measurable contributions 
to the density of states and the heat capacity: it is the thermodynamic manifestation of the topological edge structure.
Furthermore, the subdivision potential reveals phase transitions of the
edge even when they are not manifested in the bulk, thus opening a variety of 
new possibilities for investigating, manipulating, and characterizing 
topological quantum matter solely in terms of equilibrium boundary physics.
\end{abstract}

\maketitle


\section{Introduction}

Topological insulators (TI's) are phases of electronic matter protected by 
time-reversal symmetry \cite{Kane2005QSHE,Kane2005Z2,Zhang2010,Hasan2010,Zhang2011}.
Here, the topological part pertains to the presence of time-reversal conjugate
pairs of boundary states that are robust, that is, stable against perturbations 
that do not break the protecting symmetry. In TI's, the transition between 
a topologically trivial and a non-trivial phase is usually described in terms 
of band inversion, where the gap between two energy bands closes and creates 
an avoided crossing hosting the protected boundary states. This phenomenon is 
seen in HgTe quantum wells, for example, which undergo a topological phase
transition as a function of quantum well thickness \cite{Bernevig2006}. 
These edge modes have been detected through transport measurements \cite{Konig2007}.
In the three dimensional case, the boundary modes are also conveniently described spectroscopically
using ARPES \cite{Brune2011}. However, a thermodynamic description of topological boundary states is missing.

A problem one encounters is that band topology is determined for infinite, translationally invariant systems
in terms of Bloch Hamiltonians \cite{Kitaev2009,Ryu2010}, whereas the 
edge states of the system are only found in finite systems with boundaries. 
The bulk-boundary correspondence describes this connection between band 
topology and edge states. If one tries to apply the thermodynamic formalism 
to topological phases of matter, one immediately discovers that there is 
no thermodynamic bulk-boundary correspondence. The thermodynamic potentials 
only depend on the energy levels of the Hamiltonian and the density of states, 
while the topology of the bands depends on the eigenstates: two models can have 
the same spectrum in the bulk, but different topological characteristics. 
Since in conventional thermodynamics the energy is additive with respect to 
the extensive variables like entropy $S$, volume $V$, and particle number $N$,
the thermodynamic contribution of the edge states is lost in the 
conventional thermodynamic limit. 

In this work, we show that the solution to this conundrum lies in Hill's refinement of conventional 
thermodynamics \cite{Hill}. As computed from statistical mechanics, 
the thermodynamic potentials (entropy or any of the various free energies) 
do not typically scale linearly with the extensive variables of the system due to finite size and boundary effects.
In order to account for this feature within the thermodynamic framework, 
Hill collects the deviations from linear scaling in a new state function, 
the subdivision potential. To show how this works, we first give the necessary details from Hill's thermodynamics in Sec.~\ref{HT}.
We then discuss in Sec.~\ref{Gibbs} how this relates to the more traditional method of treating boundary effects due to Gibbs. Finally, in Sec.~\ref{App} we apply the developed
theory to the Bernevig-Hughes-Zhang (BHZ) model for HgTe quantum wells, to show how this works in practice. Our main results are presented in Sec.~\ref{Res}: We show that the thermodynamic density of states and the specific heat signatures of the topological edge states can be experimentally detected. Moreover, we find that the topological phase transition is accompanied by a thermodynamic phase transition on the boundary of the system that has no counterpart in conventional thermodynamics. Our conclusions are summarised in Sec.~\ref{Con}.

\section{Hill Thermodynamics}\label{HT}
Let us consider a finite system of size \(V\), in contact with an 
environment at temperature \(T\) and electronic chemical potential $\mu$.
We take the extensive variable \(V\) to be the number of sites in the 
lattice associated to a tight-binding model of a TI. Formally, 
\(V\) must be a fluctuating parameter, which can be achieved by considering a reservoir of ions capable of becoming attached 
to the lattice. Then, conjugate to $V$ there is a variable $\nu$ 
characterizing the thermodynamic response of the band structure to an 
increase in the number of lattice sites \footnote{One can also consider the dependence of the thermodynamics on $V$ if it is not a fluctuating parameter, but then one would formally be comparing different thermodynamic systems, which is experimentally more feasable, but theoretically inelegant.}.

For a system with these independent state variables, one uses the grand potential \(\Phi\). In Hill's thermodynamics, it still obeys the conventional relation  
\begin{equation}\label{Tidentity}
d\Phi=-SdT-\nu dV-Nd\mu,
\end{equation}
as in ordinary thermodynamics. Hence, the connection with the 
microscopic behaviour is made through the statistical-mechanical
partition function,
\begin{equation}\label{GibbsEnsemble}
\Phi=-k_B T \ln\left\{\mathrm{Tr}\exp\left[-(H-\mu N)/ k_B T\right]\right\},
\end{equation}
with $k_B$ denoting the Boltzmann constant and $H$ the Hamiltonian 
of the TI. However, the differential equation for \(\Phi\) does
not integrate to \(-\nu V\) by way of Euler's theorem because non-linear 
scaling with the extensive variable \(V\) is allowed within Hill's 
framework. The new thermodynamic variable \(\hat{\nu}=-\Phi/V\) determines
the subdivision potential \(X\) as 
\begin{eqnarray}
X=-(\hat{\nu}-\nu)V.
\end{eqnarray}
Hill thermodynamics was originally developed for small systems, where
highly non-linear thermodynamic potentials are computed by considering an independent ensemble of such systems \cite{Hill}.
The approach also works if the individual small systems are not independent, since it allows for a systematic computation of potentials from statistical mechanics \cite{Chamberlin1999,Chamberlin2000,Chamberlin2014}.
Interestingly, Hill thermodynamics is just as necessary for systems
as large as gravitationally bound systems \cite{Latella2015}, where 
conventional thermodynamics fails to apply because the gravitational 
force is long-ranged and universally attractive. Topologically non-trivial
states of electronic matter belong to neither one of these categories.
What makes the subdivision potential important for TI's is the 
strong dependence of the spectrum on the boundary conditions, due to the bulk-boundary correspondence. Since the boundary is small compared to the bulk, systems with a strong behavioural dependence on boundary conditions are in a sense small themselves, and hence Hill's thermodynamics is the natural framework to describe them. 

\section{Gibbs and Hill thermodynamics}\label{Gibbs}

A recurring theme in this work, is that Hill thermodynamics describes features in topological insulators that cannot be described by effective boundary theories. In order to shed further light on this statement, we now give a detailed description of the Hill formalism. We then contrast this approach to the traditional Gibbs method of effective boundary theories. Finally, we indicate where the Gibbs method breaks down in the case of topological insulators, and why.

The basic physical assumption in thermodynamics is the thermodynamic identity
\begin{align}\label{dE}
	dE=TdS-pdV+\mu dN,
\end{align}
where $T,p,\mu$ denote, respectively, temperature, pressure and chemical potential. It relates the average energy, which completely determines the system, to the thermodynamic variables. Conventional thermodynamics assumes the energy to be extensive; however, it is unable to describe the non-local behaviour of topological phase transitions. If one relaxes extensiveness, Eq.~\eqref{dE} is no longer straightforwardly integrated. Nevertheless, Hill realised that this problem may be overcome in a very astute manner: consider a macroscopic number $\mathcal{N}$ of independent copies of the system, and allow this number of copies to vary. The thermodynamic identity for the total system then reads
\begin{align}\label{HilldE}
	dE_t=TdS_t-p\mathcal{N}dV+\mu dN_t-\hat p V d\mathcal{N},
\end{align}
where the subscript $t$ stands for the total system, $V$ is the volume of an individual subsystem, and $-\hat p V$ is a formal thermodynamic response of the system to changes in $\mathcal{N}$. It is important to note that $\hat p$ might well depend on $V$ itself, but the leading term in $\hat p V$ should be linear in $V$. Now, we consider the total system at fixed $T,\mu,V$ and since $E$ must be linear in $\mathcal{N}$, using Euler's theorem we can integrate Eq.~\eqref{HilldE} to get
\begin{align}\label{HillE}
	E_t=TS_t-\hat p V\mathcal{N}+\mu N_t.
\end{align}
The energy of an individual system may be obtained by dividing Eq.~\eqref{HillE} by $\mathcal{N}.$ Using that $S=S_t/\mathcal{N}$ and $N=N_t/\mathcal{N}$, this gives $E=TS-\hat p V+\mu N$. This result holds for any $V,T,\mu$ system, and we have actually integrated Eq.~\eqref{dE} for this case. The non-extensive behaviour is naturally incorporated since $\hat p$ can depend on $V$. The deviations of Hill's thermodynamics can be naturally separated from the conventional formalism by writing
\begin{align}\label{E}
  E=TS-pV+\mu N+X.
\end{align}
Here, $X=(p-\hat p)V$ defines Hill's subdivision potential; it is an extra degree of freedom that characterises the non-extensiveness of the system.

By inserting the ansatz
\begin{equation}\label{HillPotential}
\frac{\Phi(\mu,T,W L)}{L}=\phi_0(\mu,T)+\phi(\mu,T)W
\end{equation}
into the Hill formalism, as in the main text, we are clearly separating the bulk from the boundary effects in the free energy. 

Gibbs has also developed a method to describe boundary effects in thermodynamics, which was originally used to describe classical fluids \cite{Gibbs1878}, and will also be used in this work. In 1878, Gibbs described a thermodynamic approach to surface tension relying on the hypothesis that bulk and boundary could be treated as independent systems in some approximate sense. In Gibbs' approach, the free energy of the fluid system acquires a term
proportional to a suitable power (e.g. $2/3$) of the volume. The phenomenological success of Gibbs' approach is remarkable, since there is no sharp surface associated to any actual microscopic fluid system. 

In Gibbs' approach, based on conventional thermodynamics, the area of the boundary has to be an extensive thermodynamic variable in itself. This implies that one treats the surface of the system as a separate thermodynamic system with its own energetics, independent from the bulk. Consider, for example, a bulk system $B$ with a boundary $b$. The bulk system will have energy
\begin{equation*}
U_B=S_B T_B-\nu V+N_B\mu_B,
\end{equation*}
and the boundary has its own energy
\begin{equation*}
U_b=S_b T_b- \gamma A+N_b \mu_b.
\end{equation*}
Applying the equations of thermodynamic equilibrium, $T_B=T_b=T$ and $\mu_B=\mu_b=\mu$, the total energy $U$ reads
\begin{equation*}
	U= T(S_B+S_b)-\nu V -\gamma A + \mu(N_B+N_b).
\end{equation*}
Considering the grand potential $\Phi_i=U_i-S_i T_i-\mu_i N_i$ for both subsystems, and defining the total grand potential ${\Phi=\Phi_B+\Phi_b}$, we obtain
\begin{equation}\label{Phi}
	\Phi=-\nu V - \gamma A.
\end{equation}
Because the bulk and the boundary are separate systems, with their own grand potential, the thermodynamic identity reads
\begin{equation}\label{GibbsdPhi}
d\Phi=-SdT-Nd\mu-\nu dV-\gamma dA.
\end{equation}
This is crucially different from the Hill approach: although Hill also uses Eq.~\eqref{Phi},
the thermodynamic identity is still given by Eq.~\eqref{dE}, which does not contain the $dA$ term,
so that the bulk and boundary are thermodynamically connected in a natural way. In this case, the $\gamma A$ term and possible additional terms define the subdivision potential $X$. 

It is also possible to consider the bulk and boundary as a connected system in the Gibbs approach. In that case, one writes $A(V,\mu,T)$, so that 
\begin{equation*}
dA=\frac{\partial A}{\partial T}dT+\frac{\partial A}{\partial \mu}d\mu+\frac{\partial A}{\partial V}dV.
\end{equation*}
By using this identity, Eq.~\eqref{GibbsdPhi} becomes formally equivalent to the standard thermodynamic identity. However, the starting point is fundamentally different, and writing $A(V,\mu,T)$ is tantamount to distilling a boundary theory from the total Hamiltonian. It does mean that if a sensible boundary theory can be written down, the Gibbs and Hill approaches will yield the same results. This coincidence is a powerful tool in analysing topological boundary effects. Although the subdivision potential $X$, and hence the boundary behaviour, can be obtained from $\Phi$ by looking at the different volume scalings, there is no clear way to separate finite-size effects from topological behaviour. However, there is a natural way to define a boundary theory for non-trivial topological insulators: diagonalise the Hamiltonian, and single out boundary states based on the localisation of the eigenfunctions. We will find that this effective theory is a powerful tool in interpreting our results when applicable. However, it is Hill's thermodynamics that allows one to find the regime where this is so. The above procedure for writing down an effective theory works precisely if the linear regime from Eq.~\eqref{HillPotential} has set in. As we discuss in the next section, the minimum width $W$ for which this holds is dependent on the gap size. This occurs because the edge states merge into the bulk at the phase transition, and an effective theory can only be expected to hold if the system is larger than the decay length of the edge states. Alternatively, the low energy theory becomes conformally invariant at the topological phase transition, so an ansatz of the form Eq.~\eqref{HillPotential} cannot be correct. However, for systems larger than a floating cutoff width $W_0$ depending on the gap size, we get Eq.~\eqref{HillPotential} as a generalised thermodynamic limit. This thermodynamic limit is equivalent to a Gibbs effective theory, hence, such an effective Gibbs theory can describe the low temperature thermodynamic responses as long as the gap remains large enough. However, if one wants to correctly describe the phase transition, one needs to keep the system size above the floating cutoff, which can be done naturally in the Hill thermodynamics by looking at the scaling, but not using an effective theory, since it is not clear precisely when an edge state has merged into the bulk.

\section{Application to Bernevig-Hughes-Zhang model}\label{App}

We will now apply the developed formalism to the paradigmatic Bernevig-Hughes-Zhang 
(BHZ) model of HgTe/CdTe quantum wells \cite{Bernevig2006}. The Hamiltonian on the infinite plane will decompose as
\begin{equation}\label{BlochHam}
H=\int^\oplus_{BZ} H_k,
\end{equation}
where $BZ$ stands for Brillouin zone, and the Bloch Hamiltonian reads
\begin{widetext}
\begin{equation}\label{Ham}
H_k=-A \sin(k_x)\sigma_x-A\sin(k_y)\sigma_y-\left\{M+2B[2-\cos(k_x)-\cos(k_y)]\right\}\sigma_z.
\end{equation}
\end{widetext}
Here, $k_i$ denotes momentum in the $i$ direction, $\sigma_i$ are $2\times2$ 
Pauli matrices, with $i = x,y,z$, and $A,B$ and $M$ are parameters depending 
on the thickness of the quantum well. For $M<0$, the system is in a topologically non-trivial phase, 
whereas for $M>0$ the gap is trivial.
We consider a ribbon with a finite width $W$ and a length $L=600\approx\infty$ (so that $V=LW$), and impose the corresponding boundary conditions on Eq.~\eqref{BlochHam}. Then, we calculate the grand potential $\Phi/L$ numerically according to Eq.~\eqref{GibbsEnsemble}.

The subdivision potential is extracted from the ansatz 
\begin{equation}\label{HillPotential}
\frac{\Phi(\mu,T,W L)}{L}=\phi_0(\mu,T)+\phi(\mu,T)W.
\end{equation}
Here, $\phi_0=X/L$ is essentially the subdivision potential of the BHZ model. A linear fit is then performed in $W$, to obtain $\phi_0$ and $\phi$ for the given values of $\mu$ and $T$. One can obtain $\phi$ and $\phi_0$ as a function of $\mu$ and $T$ by evaluating them on a grid, and interpolating. We let one single parameter vary, while keeping all others constant, and we use one-dimensional Hermite interpolation. Our results indicate that the above ansatz correctly describes the relevant features of the model for large $W$, but the interested reader is referred to the supplementary material for a detailed error analysis of the fitting procedure. The parameters $A=B=1$ are fixed for numerical convenience and clarity in the results. From Eq.~\eqref{HillPotential}, it is readily derived that $\nu=-\phi$ and $\hat \nu=-\phi-\phi_0/W$, which indeed become equal as $W\rightarrow \infty$, as expected for large systems. However, we will demonstrate that $\phi_0$ cannot be neglected for topologically non-trivial systems, showing that Eq.~\eqref{HillPotential} defines a generalised thermodynamic limit, appropriate for topological insulators with boundary.

\section{Results}\label{Res}

\begin{figure*}
\includegraphics[width=\linewidth]{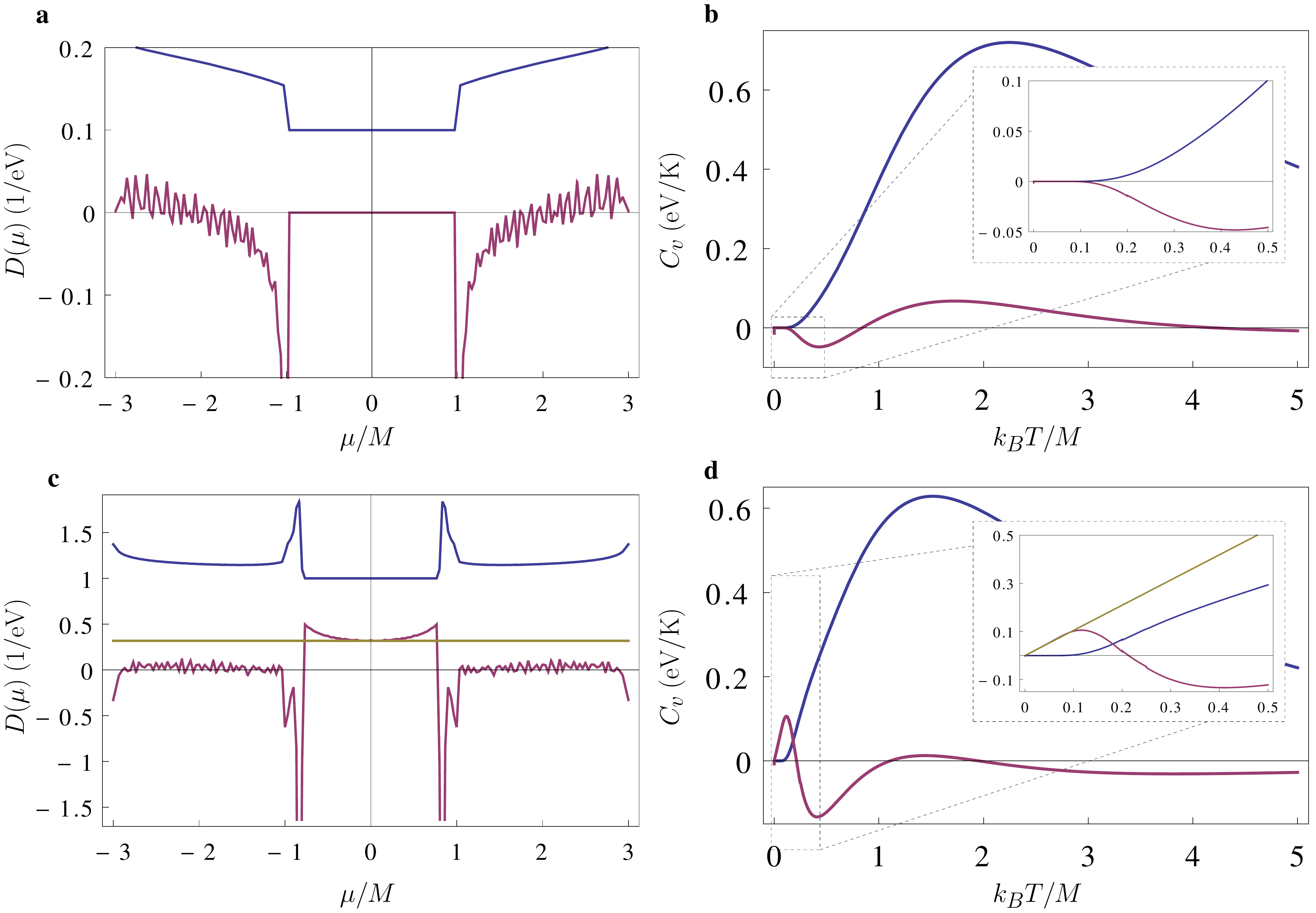}\\
\caption{\label{F1}(a) In blue, $D^{B}(\mu)/W$ is shown for $M=1$; it has been shifted up by $0.1$ for greater visibility. As expected, it vanishes in the energy gap. In red, $D^{b}(\mu)$ for the same parameters, which is the DOS on the edge. Since the system is not in a topological phase, the DOS at the edge also vanishes in the gap. (b) In blue, $C^B_v(T)/W$ is shown as a function of $T$ for $M=1, \mu=0$. As expected, it vanishes to linear order as $T\rightarrow 0$. In red, $C^b_v(T)$ for the same parameters. Since the system is not in a topological phase, $C^b_v$ also vanishes to linear order as $T\rightarrow 0$. (c) The same as in (a), except that the blue curve has been shifted upwards by $1$, and the system is in the topological phase $M=-1$. Furthermore, the line $1/\pi$ has been added in yellow to emphasise that the edge has a non-vanishing DOS in the gap. (d) The same as in (b), but in the topological phase $M=-1$. One sees that $C^b_v$ now is linear at the edge for low $T$. In the inset, the straight line with slope $\pi/3$ has been added to emphasise the linear scaling of $C^b_v$ with temperature.}
\end{figure*}

Using the thermodynamic identity, various thermodynamic responses can be calculated, and due to Eq.~\eqref{HillPotential}, these naturally split into a boundary and a bulk contribution. In Fig.~\ref{F1}(a) and Fig.~\ref{F1}(c) (Fig.~\ref{F1}(b) and Fig.~\ref{F1}(d)), we plot the $T=0$ ($\mu=0$) density of states (heat capacity at constant volume) for the bulk (B) ${D^B:=-\partial^2\phi/\partial\mu^2}$ (${C^B_v:=-T \partial^2\phi/\partial T^2}$) , and for the boundary (b) ${D^b:=-\partial^2\phi_0/\partial\mu^2}$ (${C^b_v:=-T \partial^2\phi_0/\partial T^2}$)  in blue and in red, respectively. In Fig.~\ref{F1}(a) (Fig.~\ref{F1}(b)), the system is in the trivial phase $M=1$, whereas in Fig.~\ref{F1}(c) (Fig.~\ref{F1}(d)), the system is the topological phase $M=-1$.

Our data show clearly that $D^B$ vanishes in the energy gap $|\mu|<1$ both 
in the topological and in the trivial phase. However, due to the presence of edges, there is also a non-vanishing contribution $D^b$. To interpret this contribution, it is important to note that $\phi_0$ contains not just the topological edge states, but also other finite size effects, as evidenced by the finite value of $\phi_0$ even if $M>0$. Outside the gap, non-topological finite size effects are dominated by the discreteness of the spectrum, and $D^b$ is essentially just noise. The noise becomes reduced in magnitude for non-zero temperatures, which smooths 
out the discreteness of the spectrum, or by fitting Eq.~(\ref{HillPotential}) to 
a larger range of $W$ values. 

On the other hand, the energy spectrum of the edge states is much less dependent on the width $W$, and not subject to noise. Inside the gap, $D^B=0$ and the scaling of the density of states with $W$ vanishes, making $D^b$ the only term for all system sizes. In this case, it can be interpreted as the density of topological edge states, since finite-size effects only show up in combination with bulk behaviour. Indeed, the Dirac states at the edge disperse with $E=A k+\mathcal{O}(k^3)$ \cite{Hasan2010}, yielding a DOS of $1/(A\pi)$ at $\mu=0$ according to the Gibbs approach; the line $1/\pi$ (obtained by putting $A=1$) has been added in yellow in Fig.~\ref{F1}(c). 

To first order in \(T\), the heat capacity behaves as
\begin{equation}\label{HeatCap}
C_v=\frac{\pi}{3} D k_B T
\end{equation}
since a Dirac fermion has conformal charge $1$ \cite{DiFrancesco}. Hence, to linear order the bulk $C^B_v$ necessarily vanishes at low temperatures for $\mu=0$.  However, while we expect the boundary $C^b_v$ to vanish at low temperatures in the trivial phase at $M=1$ (Fig.~\ref{F1}(b)), for the topological phase with $M=-1$ the specific heat amounts to ${C_v^b=\pi k_B T/3}$, which can be obtained by simply substituting $D(0)=1/\pi$ into Eq.~(\ref{HeatCap}). Indeed, in Fig.~\ref{F1}(d), we observe a linear scaling of $C^b_v$ at low temperatures. In the inset, the yellow line with slope $\pi/3$ has been added to confirm that the low temperature heat capacity derives from the edge states.

\begin{figure*}
\includegraphics[width=\linewidth]{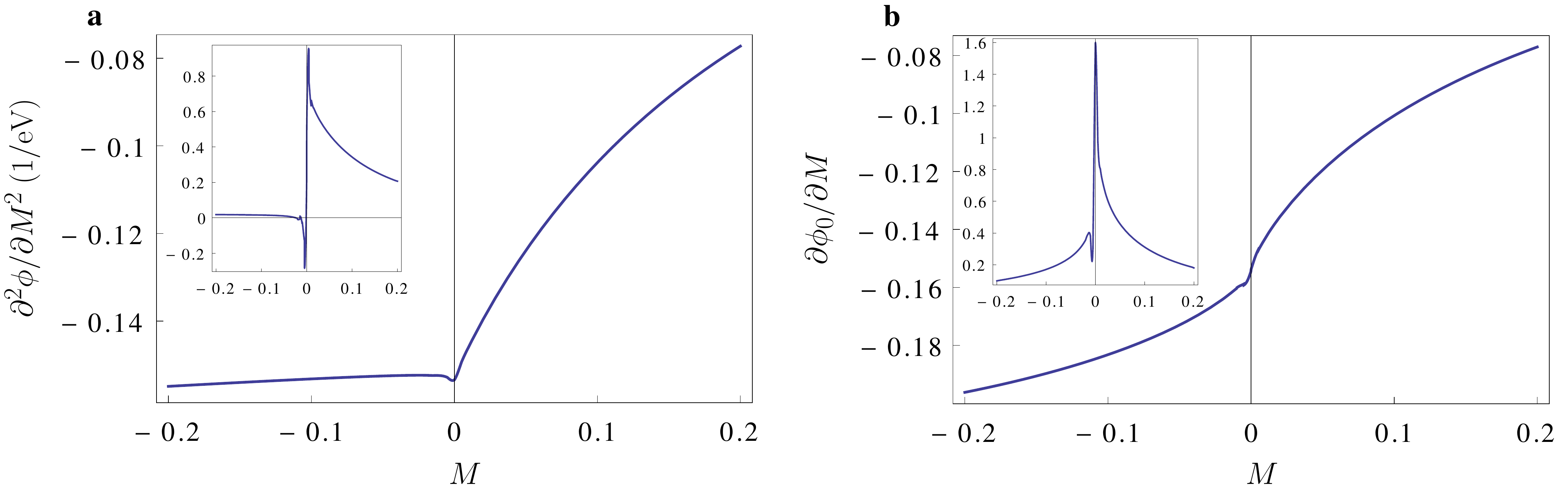}\\
\caption{\label{F2}(a) The second derivative $\partial^2\phi/\partial M^2$ is shown as a function of $M$. A kink is visible at $M=0$, indicating that the bulk undergoes a third-order phase transition (see the inlay, where the discontinuity in the third derivative is depicted). (b) The first derivative $\partial\phi_0/\partial M$ is shown as a function of $M$. A kink is visible at $M=0$ indicating that the edge undergoes a second-order phase transition (the discontinuity in the second derivative is depicted in the inlay), which can be considered the order of the topological phase transition.}
\end{figure*}

In both of these cases, the $\phi_0$ term in Eq.~(\ref{HillPotential}) could not be dropped because a derivative of $\phi$ vanished, and hence the contribution from $\phi_0$ became dominant. This shows that for edge effects to become irrelevant, it is not only required that $\hat \nu\rightarrow \nu$, but that this also holds for all derivatives. There is another situation where the derivatives of $\hat \nu$ fail to converge, which is if a phase transition occurs at the boundary. In Fig.~\ref{F2}, we depict the behaviour of the thermodynamic potential in terms of the parameter $M$ that controls the topological phase transitions in the BHZ model. In Fig.~\ref{F2}(a), $\partial^2\phi/\partial M^2$ is shown, which exhibits a slightly smoothed kink at $M=0$. This smoothing is precisely as large as the sample spacing in $M$ from which we interpolated to obtain the graph, indicating that it is a numerical artefact. The inlay shows $\partial^3\phi/\partial M^3$, which is discontinuous at $M=0$.  Calculating the same graph for the system on an infinite plane yields the same behaviour, except that the kink is sharper. Therefore, the closing of the band gap is detected by the bulk free energy and characterised as a third-order phase transition. It is reminiscent of a lambda phase transition, and the divergence occurs in the non-topological regime.
In contrast, Fig.~\ref{F2}(b) shows that $\partial\phi_0/\partial M$ exhibits a kink. The inlay shows that the transition is again of lambda type, with the divergence occurring in the non-topological regime, but this time the order is different: the edge undergoes a second-order phase transition. This phase transition cannot be described by an effective boundary theory through the Gibbs method, as can be seen because the divergence is at the trivial side, where there are no edge states. This occurs because the phase transition is driven by the merging of the topological edge states into the bulk. The precise moment when an edge state ceases to be an edge state cannot be determined from its localisation. However, any contribution to the free energy coming from topological states should scale with the edge length of the system. As such, even if we do not precisely know which states are edge states, $\phi_0$ naturally captures how many there are in total. Therefore, it is not the presence of the edge states that determines the boundary scaling, but the boundary scaling that determines the edge states.


\begin{figure}
\includegraphics[width=\linewidth]{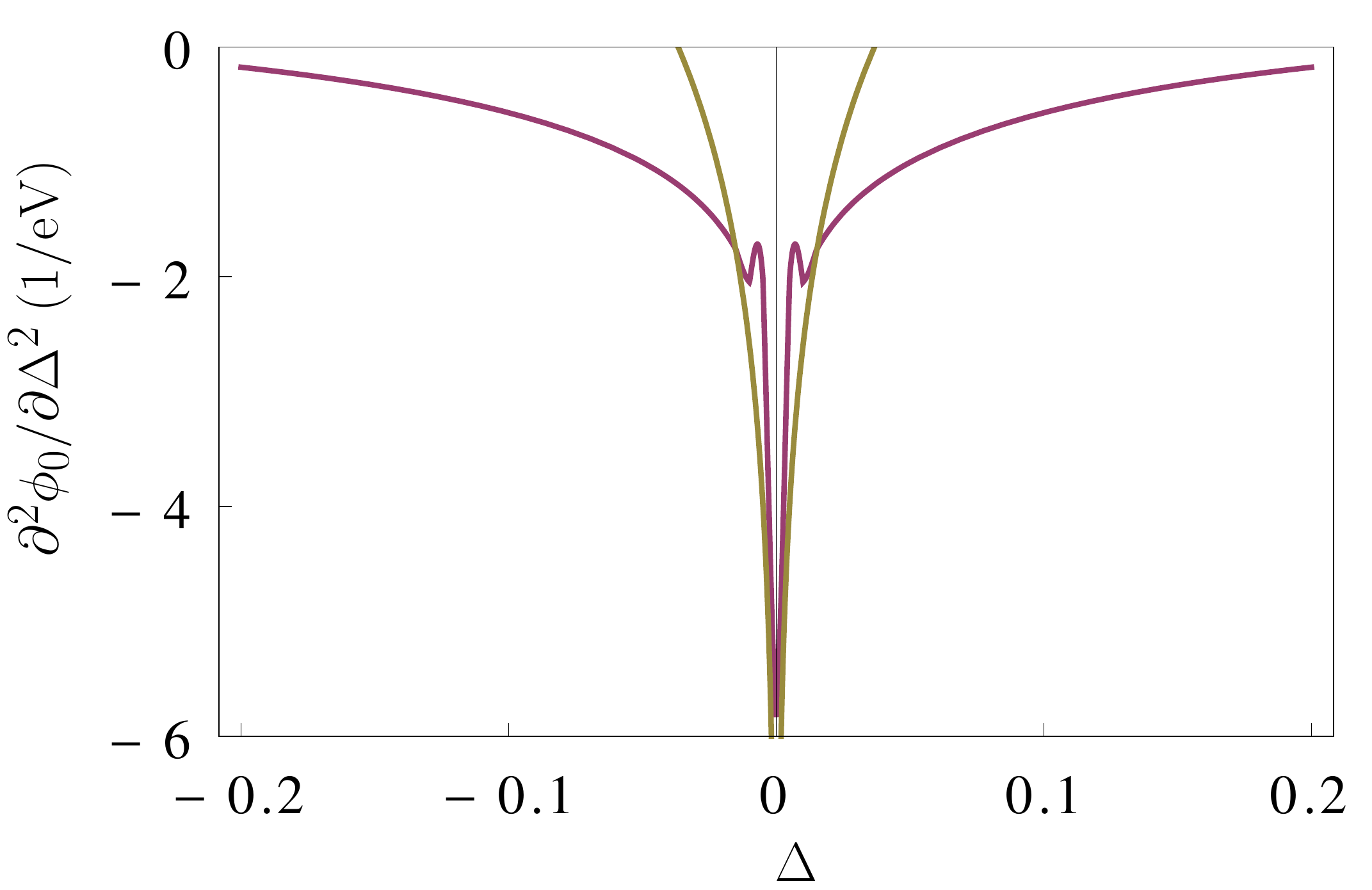}\\
\caption{\label{F3}In red, $\partial^2\phi_0/\partial \Delta^2$ is shown as a function of $\Delta$. The smoothing of an infinite peak is visible, which indicates the presence of a third-order phase transition at the boundary of the system. In yellow, $\partial^2\phi_e/\partial \Delta^2+6$ is shown, using Eq.~(\ref{DiracPotential}). The agreement between the two curves indicates that the phase transition at the edge of the model comes from the opening of a gap for the edge states.}
\end{figure}

The absence of a sensible boundary theory might cause doubt whether $\phi_0$ truly detects a thermodynamic phase transition related to the
appearance of topological edge states, rather than an echo of the bulk phase transition. This issue can be clarified by adding an on-site superconducting pairing to the BHZ-model. The same effect and results would be obtained 
from induced superconductivity on the edge \cite{Hart2014}. The Bloch Hamiltonian for this 
system is 
\begin{equation}\label{SCHam}
H_{\Delta,k}:=\left(\begin{array}{cc}
H_k &\Delta \\
\Delta & -H_k^*
\end{array}\right),
\end{equation}
where $H_k$ is the Hamiltonian from Eq.~(\ref{Ham}) and $\Delta$ is the
superconducting pairing parameter. Adding superconducting pairing has added particle-hole symmetry to the system;
the bulk stays gapped, but now the gapless edge states acquire a mass 
$\Delta$ \cite{EzawaGap2013}.
Since the gap does not close, $\phi$ is a smooth function of $\Delta$, and 
no phase transition occurs in the bulk. However, the subdivision potential 
$\phi_0$ detects the opening of the mass gap for the Dirac electrons at the 
edge as a continuous boundary phase transition. This can be seen in Fig.~\ref{F3}, 
where $\partial^2\phi_0/\partial \Delta^2$ is shown in red. A clear divergence 
is present in $\partial^2\phi_0/\partial \Delta^2$ at $\Delta=0$, where the particle-hole symmetry is broken.

Because adding a Cooper pairing does not merge edge states into the bulk, but only gives them a mass, it is possible to describe a phase transition in $\Delta$ using an effective boundary theory in the manner of Gibbs, and thereby confirm our interpretation. This theory reads
\begin{equation}\label{DiracHam}
H_{e,k}:=k \sigma_x +\Delta \sigma_z,
\end{equation}
with a momentum cutoff $|k|<1$ (since $A=B=-M=1$, the edge states exist only for this range of $k$ values). The free energy 
per unit length of $H_{e,k}$ is  
\begin{equation}\label{DiracPotential}
\phi_e=\Delta^2\ln\left(\frac{1+\sqrt{1+\Delta^2}}{\Delta}\right)+\sqrt{1+\Delta^2}.
\end{equation}
The corresponding value of $\partial^2\phi_e/\partial \Delta^2$ is shown in Fig.~\ref{F3} in yellow. The divergence at $\Delta=0$ indicates a continuous phase transition due to the appearance of gapless edge modes in Eq.~(\ref{DiracHam}). The behaviour of the subdivision potential $\phi_0$ is compatible with that of $\phi_e$, indicating that $\phi_0$ truly detects topological edge behaviour, since it gives the same qualitative results as an effective boundary theory.


\section{Conclusions}\label{Con}

Our general scheme for extracting thermodynamic signatures of the 
bulk-boundary correspondence, illustrated here for the paradigmatic BHZ model 
of the quantum-spin Hall effect in HgTe/CdTe quantum wells, comprises a novel 
tool kit for detecting these elusive states of matter in terms of equilibrium
measurements. The scheme is based on the idea that for topologically non-trivial
systems, a non-conventional thermodynamic limit exists: even though 
the system becomes infinitely large, the boundary is always present. The 
boundary is taken into account as in Eq.~(\ref{HillPotential}), in terms of the subdivision potential at the heart of Hill thermodynamics. The subdivision potential captures thermodynamic signatures unique to 
topological insulators, including a DOS in the bandgap observable in
transport measurements \cite{Konig2007}, and a contribution to 
the electronic heat capacity, linear for the BHZ model. Most notably, this
contribution might be relevant for identifying topological phases in 
ultracold-atom systems, where transport experiments are challenging
and the absence of phonons makes the measurement of fermionic
heat capacity easier. 

Although the edge contributions to the density of states and the specific heat can be captured by an effective model that artificially separates the bulk and the boundary contributions, the same does not hold when describing generic topological phase transitions.
If a system undergoes a phase 
transition within the same symmetry class, the bandgap necessarily closes 
and a bulk phase transition takes place. The appearance/disappearance of
topologically protected edge states gives rise to an accompanying boundary 
phase transition, which can only be classified by the subdivision potential, and does not need to be of the same order as the one in the bulk.
Furthermore, one also observes phase transitions between different symmetry classes, which occur without a 
closing of the bulk gap. In this case, the topological phase transition 
occurs solely at the boundary. The fact that $\phi_0$ detects a phase 
transition even if the bulk gap does not close shows that the subdivision
potential describes the edge behaviour of the system. This makes it a quantity of prime interest for investigating topological phase transitions, and allows for a classification of their order within the well known Ehrenfest scheme. These results open a new set 
of possibilities for experimentally detecting topological order by delicate
but standard thermodynamic methods, and provide a deeper understanding of 
the effect of edges in the abstruse field of topological insulators.

\begin{acknowledgements}
The authors would like to acknowledge A. Bernevig and L. Molenkamp for useful discussions. This work is part of the D-ITP consortium, a program of the Netherlands Organisation for Scientific Research (NWO) that is funded by the Dutch Ministry of Education, Culture and Science (OCW).
\end{acknowledgements}

\appendix

\section*{Appendix: Fitting analysis}
Key to our work is the assumption that the grand potential $\Phi$ has the asymptotic form given in Eq.~(\ref{HillPotential}).
Since numerical calculations are necessarily done for finite samples, the question rises for which values of $W$ deviations from the asymptotic behaviour in Eq.~(\ref{HillPotential}) become negligible. Here, we provide a detailed description of the way these deviations depend on the various parameters in the system. The deviation, as a function of $W$, depends on $\mu$ and $T$, and also on the gap size. Throughout this appendix, as well as in the main text, $A=B=1.$ We will vary the parameters $\mu$, $T$, and $M$, and study the relative error $(\Phi_W-\Phi)/\Phi$, where $\Phi_W$ is given by Eq.~(\ref{GibbsEnsemble}) for sample width $W$, and $\Phi$ is the corresponding value after fitting to Eq.~(\ref{HillPotential}). The range of $W$ along which we fitted to obtain $\Phi$ will be mentioned for each specific case.

\begin{figure*}[htb]
\includegraphics[width=\linewidth]{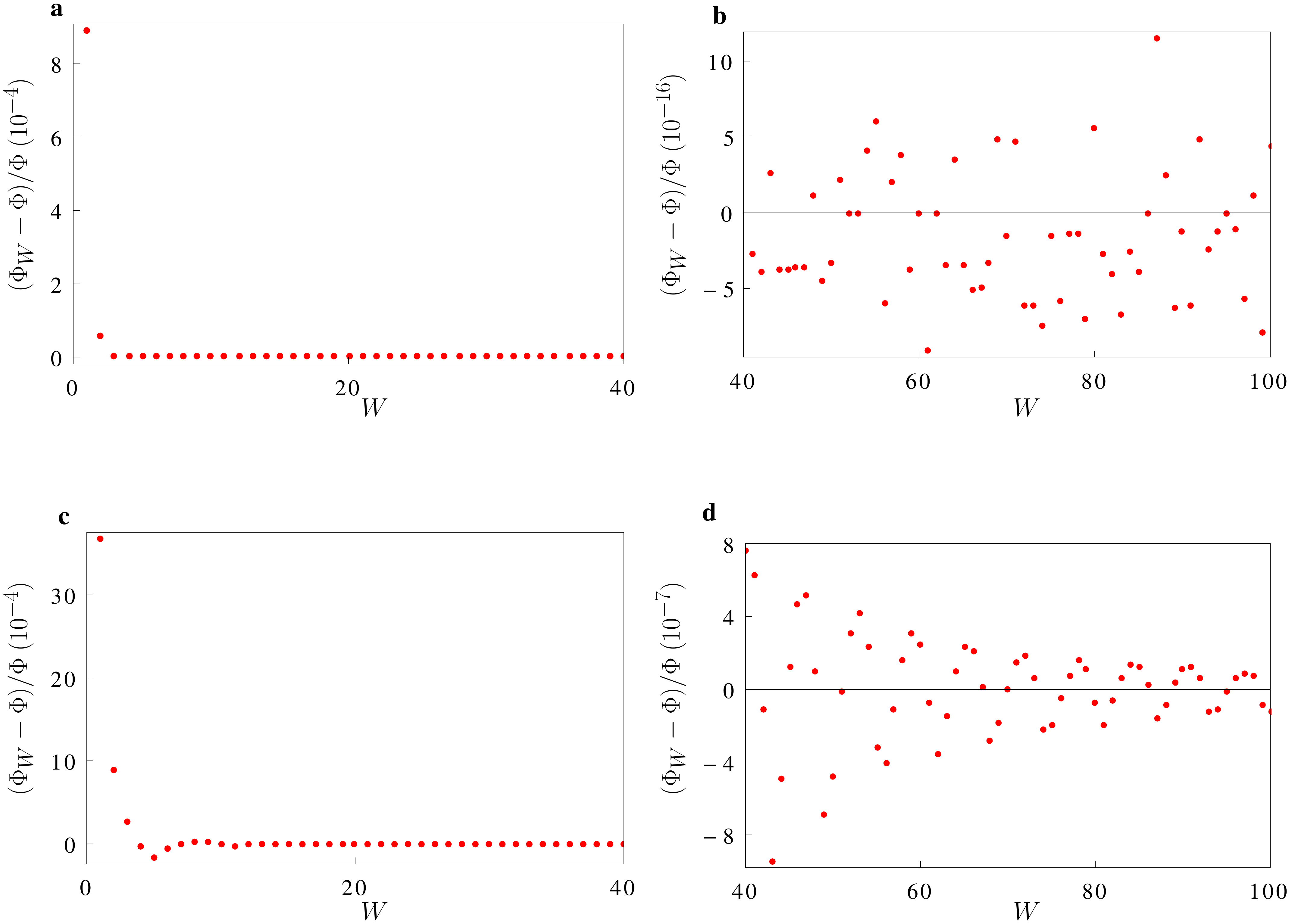}
\caption{\label{E1} (a) The relative error between $\Phi_W$ and $\Phi$, for $\mu=T=0$ and $M=-1$. We have fitted along the interval $40\leq W\leq 100.$ (b) Same as in (a) but the range of $W$ has been changed. (c) Same as in (a) but for $\mu=-4/3$. (d) Same as in (b) but for $\mu=-4/3$.}
\end{figure*}

In Fig.~\ref{E1}, the relative error has been plotted for the trivial phase $M=1$, at $T=0$, for small values of $W$, although the fitting was done for $40\leq W\leq 100,$ where the linear behaviour in $W$ has set in. In this way, one observes the small $W$ deviations from Eq.~\ref{HillPotential} without including them in the fit. The results are shown in Fig.~\ref{E1}(a) and (b) for $\mu=0$, i.e. in the gap. Similarly, the results are shown in Fig.~\ref{E1}(c) and (d) for $\mu=-4/3$, i.e. in one of the bulk energy bands. We see that for very small system width, there is a large deviation from the linear relation in Eq.~(\ref{HillPotential}), but the error quickly decreases. For $\mu$ in the gap, the error quickly becomes negligible, while for $\mu$ in one of the energy bands, the error is much larger , but also shows a clear structure (notice the unequal scales $10^{-16}$ in Fig.~\ref{E1}(b) and $10^{-7}$ in Fig.~\ref{E1}(d)). A likely cause of this error is the discreteness of the spectrum, which causes $\Phi_W$ to deviate from $\Phi$ as the system width is varied. This deviation occurs because varying the system width causes individual energies jump from above to below the chemical potential and vice versa. In the gap there are no states near the chemical potential, so this effect is absent, significantly reducing the error. 

\begin{figure}[htb]
\begin{center}
\includegraphics[width=\linewidth]{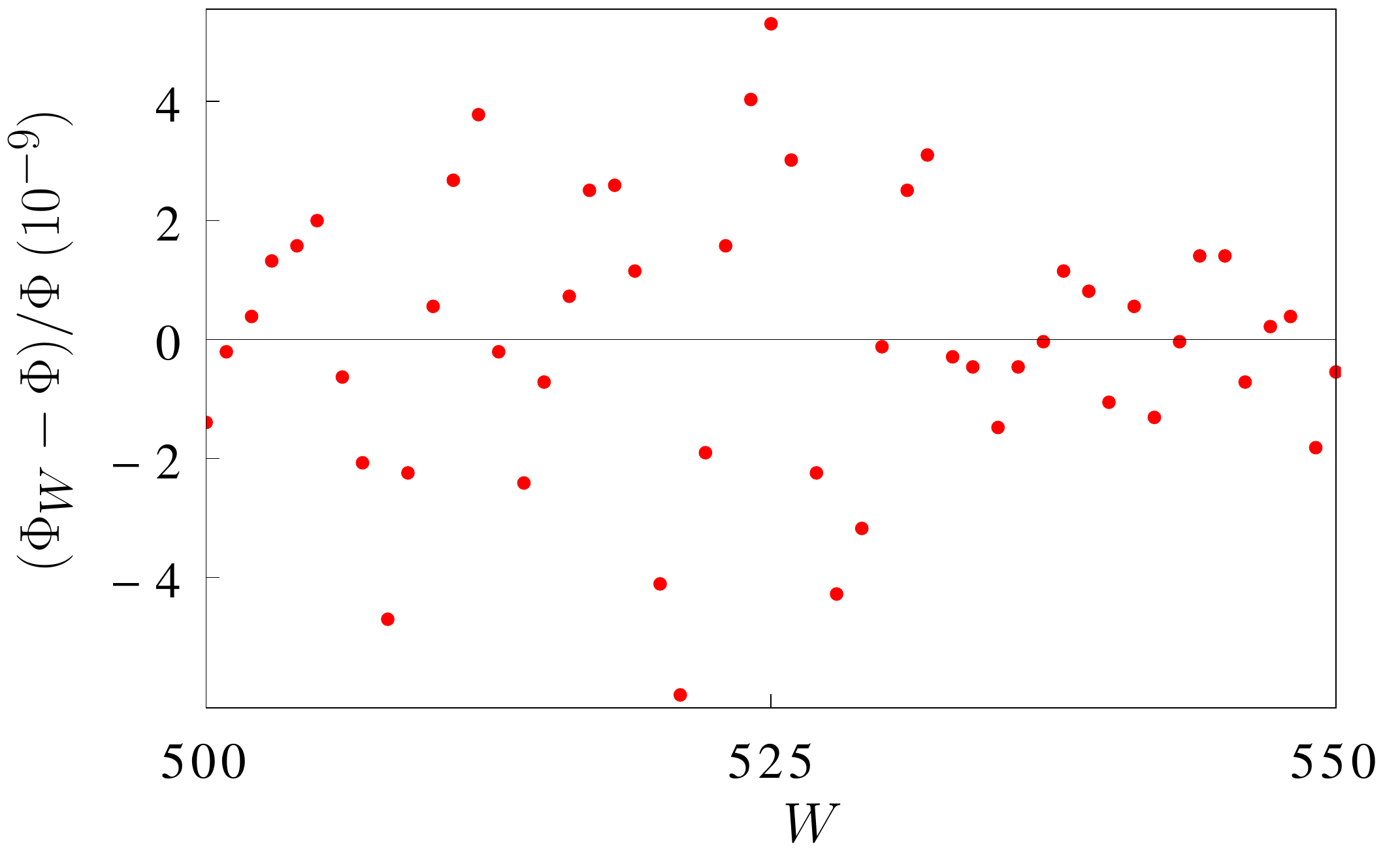}
\end{center}
\caption{\label{E2} The relative error between $\Phi_W$ and $\Phi$, for $T=0$, $\mu=-4/3$, and $M=-1$. We have fitted along the interval $500\leq W\leq 550.$}
\end{figure}

In Fig.~\ref{E2}, the relative error is plotted for $\mu=-4/3$ at $T=0$ and $M=1$ for a fit along $500\leq W\leq 550.$ The relative error has decreased by two orders of magnitude compared to Fig.~\ref{E1}(d), which means that the absolute error has decreased by one order of magnitude. This implies that the energy spectrum becomes less dependent on $W$ as $W$ increases. Furthermore, the behaviour of the relative error is qualitatively similar if one takes $M=-1$, which puts the system in the topological phase. This shows that the energy spectrum of the edge states is highly independent of $W$ even at small width, as can be expected from their strong localisation on the edge.

\begin{figure*}[htb]
\includegraphics[width=\linewidth]{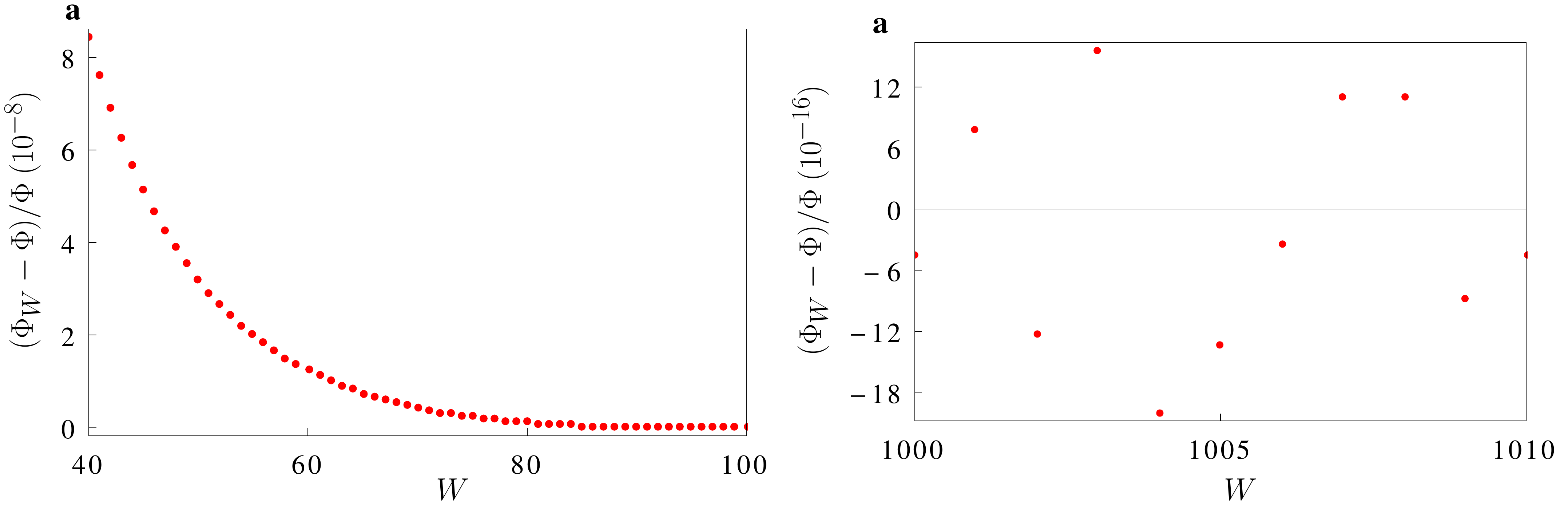}
\caption{\label{E3} (a) The relative error between $\Phi_W$ and $\Phi$, for $T=0$, $\mu=0$ and $M=0$. We have fitted along the interval $90\leq W\leq 100.$ It can be seen that the linear regime has not set in for this system size. (b) The same as in (a), but for $1000\leq W\leq 1010.$ The linear regime has set in, and the error is negligible.}
\end{figure*}

When the gap becomes much smaller, the conclusions above still hold, but the width $W$ for which the error becomes negligible is larger. In Fig.~\ref{E3}(a), the relative error has been plotted for $\mu=T=0$ at $M=0$, for a fit along the values $90\leq W\leq 100.$ The errors show that the linear regime from Eq.~(\ref{HillPotential}) has not set in yet. In contrast, Fig.~\ref{E3}(b) shows a fit along $1000\leq W\leq 1010,$ for the same parameters, and here a linear scaling clearly holds.

\begin{figure*}[htb]
\includegraphics[width=\linewidth]{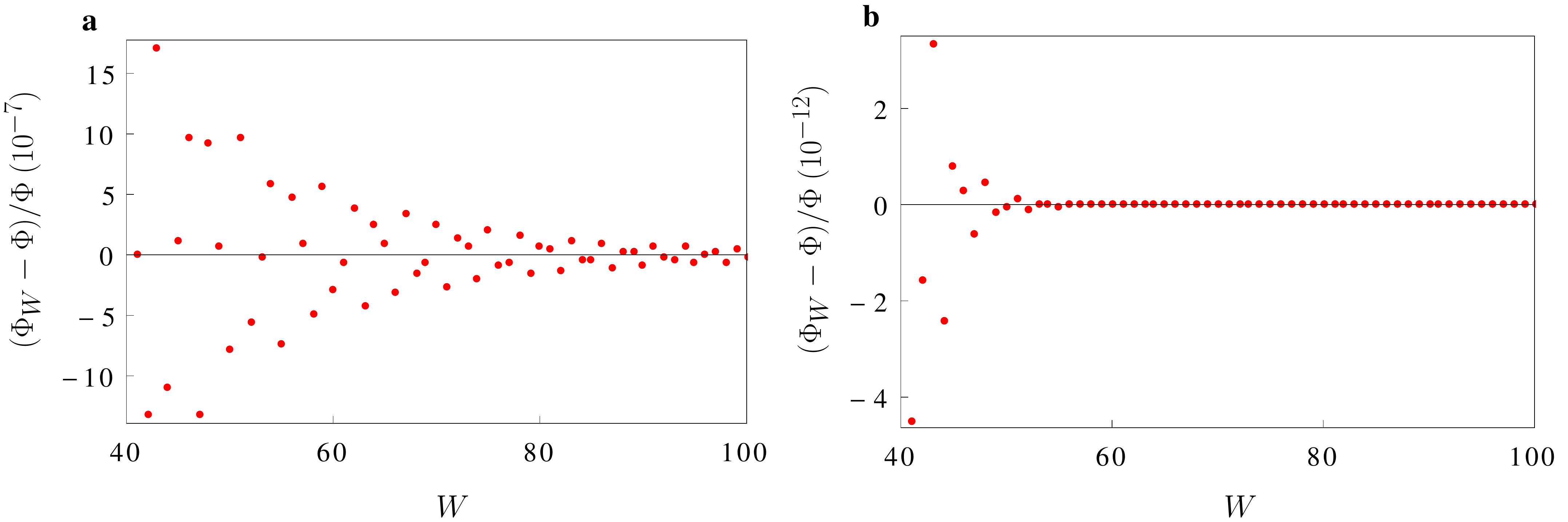}
\caption{\label{E4} (a) The relative error between $\Phi_W$ and $\Phi$, for $k_B T=1/100$, $\mu=-2$ and $M=1$. We have fitted along the interval $50\leq W\leq 100.$ The errors are of the same order as in Fig.~\ref{E1}(d). (b) The same as in (a), but for $k_B T=1/10.$ The errors are significantly smaller, since the temperature effectively smooths out the energy spectrum.}
\end{figure*}

Finally, in Fig.~\ref{E4}, the relative error is shown for $M=-1$, $\mu=-2$ for a fitting interval of $50\leq W\leq 100$. In Fig.~\ref{E4}(a), where $T=1/100$, the relative error is of the same order of magnitude as in Fig.~\ref{E1}(d), indicating that the temperature is not yet high enough to suppress the fluctuations. In Fig.~\ref{E4}(b), where $T=1/10$, the relative error has decreased by orders of magnitude, indicating that the smoothing of the Fermi-Dirac distribution at these temperatures suppresses the fluctuations in the energy spectrum.
\bibliographystyle{apsrev4-1}

\begin{thebibliography}{20}%
\makeatletter
\providecommand \@ifxundefined [1]{%
 \@ifx{#1\undefined}
}%
\providecommand \@ifnum [1]{%
 \ifnum #1\expandafter \@firstoftwo
 \else \expandafter \@secondoftwo
 \fi
}%
\providecommand \@ifx [1]{%
 \ifx #1\expandafter \@firstoftwo
 \else \expandafter \@secondoftwo
 \fi
}%
\providecommand \natexlab [1]{#1}%
\providecommand \enquote  [1]{``#1''}%
\providecommand \bibnamefont  [1]{#1}%
\providecommand \bibfnamefont [1]{#1}%
\providecommand \citenamefont [1]{#1}%
\providecommand \href@noop [0]{\@secondoftwo}%
\providecommand \href [0]{\begingroup \@sanitize@url \@href}%
\providecommand \@href[1]{\@@startlink{#1}\@@href}%
\providecommand \@@href[1]{\endgroup#1\@@endlink}%
\providecommand \@sanitize@url [0]{\catcode `\\12\catcode `\$12\catcode
  `\&12\catcode `\#12\catcode `\^12\catcode `\_12\catcode `\%12\relax}%
\providecommand \@@startlink[1]{}%
\providecommand \@@endlink[0]{}%
\providecommand \url  [0]{\begingroup\@sanitize@url \@url }%
\providecommand \@url [1]{\endgroup\@href {#1}{\urlprefix }}%
\providecommand \urlprefix  [0]{URL }%
\providecommand \Eprint [0]{\href }%
\providecommand \doibase [0]{http://dx.doi.org/}%
\providecommand \selectlanguage [0]{\@gobble}%
\providecommand \bibinfo  [0]{\@secondoftwo}%
\providecommand \bibfield  [0]{\@secondoftwo}%
\providecommand \translation [1]{[#1]}%
\providecommand \BibitemOpen [0]{}%
\providecommand \bibitemStop [0]{}%
\providecommand \bibitemNoStop [0]{.\EOS\space}%
\providecommand \EOS [0]{\spacefactor3000\relax}%
\providecommand \BibitemShut  [1]{\csname bibitem#1\endcsname}%
\let\auto@bib@innerbib\@empty
\bibitem [{\citenamefont {Kane}\ and\ \citenamefont
  {Mele}(2005{\natexlab{a}})}]{Kane2005QSHE}%
  \BibitemOpen
  \bibfield  {author} {\bibinfo {author} {\bibfnamefont {C.}~\bibnamefont
  {Kane}}\ and\ \bibinfo {author} {\bibfnamefont {E.}~\bibnamefont {Mele}},\
  }\href@noop {} {\bibfield  {journal} {\bibinfo  {journal} {Phys. Rev. Lett.}\
  }\textbf {\bibinfo {volume} {95}},\ \bibinfo {pages} {226801} (\bibinfo
  {year} {2005}{\natexlab{a}})}\BibitemShut {NoStop}%
\bibitem [{\citenamefont {Kane}\ and\ \citenamefont
  {Mele}(2005{\natexlab{b}})}]{Kane2005Z2}%
  \BibitemOpen
  \bibfield  {author} {\bibinfo {author} {\bibfnamefont {C.}~\bibnamefont
  {Kane}}\ and\ \bibinfo {author} {\bibfnamefont {E.}~\bibnamefont {Mele}},\
  }\href@noop {} {\bibfield  {journal} {\bibinfo  {journal} {Phys. Rev. Lett.}\
  }\textbf {\bibinfo {volume} {95}},\ \bibinfo {pages} {146802} (\bibinfo
  {year} {2005}{\natexlab{b}})}\BibitemShut {NoStop}%
\bibitem [{\citenamefont {{Qi}}\ and\ \citenamefont
  {{Zhang}}(2010)}]{Zhang2010}%
  \BibitemOpen
  \bibfield  {author} {\bibinfo {author} {\bibfnamefont {X.-L.}\ \bibnamefont
  {{Qi}}}\ and\ \bibinfo {author} {\bibfnamefont {S.-C.}\ \bibnamefont
  {{Zhang}}},\ }\href {\doibase 10.1063/1.3293411} {\bibfield  {journal}
  {\bibinfo  {journal} {Phys. Today}\ }\textbf {\bibinfo {volume} {63}},\
  \bibinfo {pages} {33} (\bibinfo {year} {2010})}\BibitemShut {NoStop}%
\bibitem [{\citenamefont {Hasan}\ and\ \citenamefont {Kane}(2010)}]{Hasan2010}%
  \BibitemOpen
  \bibfield  {author} {\bibinfo {author} {\bibfnamefont {M.~Z.}\ \bibnamefont
  {Hasan}}\ and\ \bibinfo {author} {\bibfnamefont {C.~L.}\ \bibnamefont
  {Kane}},\ }\href {\doibase 10.1103/RevModPhys.82.3045} {\bibfield  {journal}
  {\bibinfo  {journal} {Rev. Mod. Phys.}\ }\textbf {\bibinfo {volume} {82}},\
  \bibinfo {pages} {3045} (\bibinfo {year} {2010})}\BibitemShut {NoStop}%
\bibitem [{\citenamefont {{Qi}}\ and\ \citenamefont
  {{Zhang}}(2011)}]{Zhang2011}%
  \BibitemOpen
  \bibfield  {author} {\bibinfo {author} {\bibfnamefont {X.-L.}\ \bibnamefont
  {{Qi}}}\ and\ \bibinfo {author} {\bibfnamefont {S.-C.}\ \bibnamefont
  {{Zhang}}},\ }\href {\doibase 10.1103/RevModPhys.83.1057} {\bibfield
  {journal} {\bibinfo  {journal} {Reviews of Modern Physics}\ }\textbf
  {\bibinfo {volume} {83}},\ \bibinfo {pages} {1057} (\bibinfo {year}
  {2011})},\ \Eprint {http://arxiv.org/abs/1008.2026} {arXiv:1008.2026
  [cond-mat.mes-hall]} \BibitemShut {NoStop}%
\bibitem [{\citenamefont {Bernevig}\ \emph {et~al.}(2006)\citenamefont
  {Bernevig}, \citenamefont {Hughes},\ and\ \citenamefont
  {Zhang}}]{Bernevig2006}%
  \BibitemOpen
  \bibfield  {author} {\bibinfo {author} {\bibfnamefont {B.}~\bibnamefont
  {Bernevig}}, \bibinfo {author} {\bibfnamefont {T.}~\bibnamefont {Hughes}}, \
  and\ \bibinfo {author} {\bibfnamefont {S.-C.}\ \bibnamefont {Zhang}},\
  }\href@noop {} {\bibfield  {journal} {\bibinfo  {journal} {Science}\ }\textbf
  {\bibinfo {volume} {314}},\ \bibinfo {pages} {1757} (\bibinfo {year}
  {2006})}\BibitemShut {NoStop}%
\bibitem [{\citenamefont {{K{\"o}nig}}\ \emph {et~al.}(2007)\citenamefont
  {{K{\"o}nig}}, \citenamefont {{Wiedmann}}, \citenamefont {{Br{\"u}ne}},
  \citenamefont {{Roth}}, \citenamefont {{Buhmann}}, \citenamefont
  {{Molenkamp}}, \citenamefont {{Qi}},\ and\ \citenamefont
  {{Zhang}}}]{Konig2007}%
  \BibitemOpen
  \bibfield  {author} {\bibinfo {author} {\bibfnamefont {M.}~\bibnamefont
  {{K{\"o}nig}}}, \bibinfo {author} {\bibfnamefont {S.}~\bibnamefont
  {{Wiedmann}}}, \bibinfo {author} {\bibfnamefont {C.}~\bibnamefont
  {{Br{\"u}ne}}}, \bibinfo {author} {\bibfnamefont {A.}~\bibnamefont {{Roth}}},
  \bibinfo {author} {\bibfnamefont {H.}~\bibnamefont {{Buhmann}}}, \bibinfo
  {author} {\bibfnamefont {L.~W.}\ \bibnamefont {{Molenkamp}}}, \bibinfo
  {author} {\bibfnamefont {X.-L.}\ \bibnamefont {{Qi}}}, \ and\ \bibinfo
  {author} {\bibfnamefont {S.-C.}\ \bibnamefont {{Zhang}}},\ }\href {\doibase
  10.1126/science.1148047} {\bibfield  {journal} {\bibinfo  {journal}
  {Science}\ }\textbf {\bibinfo {volume} {318}},\ \bibinfo {pages} {766}
  (\bibinfo {year} {2007})}\BibitemShut {NoStop}%
\bibitem [{\citenamefont {Br\"une}\ \emph {et~al.}(2011)\citenamefont
  {Br\"une}, \citenamefont {Liu}, \citenamefont {Novik}, \citenamefont
  {Hankiewicz}, \citenamefont {Buhmann}, \citenamefont {Chen}, \citenamefont
  {Qi}, \citenamefont {Shen}, \citenamefont {Zhang},\ and\ \citenamefont
  {Molenkamp}}]{Brune2011}%
  \BibitemOpen
  \bibfield  {author} {\bibinfo {author} {\bibfnamefont {C.}~\bibnamefont
  {Br\"une}}, \bibinfo {author} {\bibfnamefont {C.~X.}\ \bibnamefont {Liu}},
  \bibinfo {author} {\bibfnamefont {E.~G.}\ \bibnamefont {Novik}}, \bibinfo
  {author} {\bibfnamefont {E.~M.}\ \bibnamefont {Hankiewicz}}, \bibinfo
  {author} {\bibfnamefont {H.}~\bibnamefont {Buhmann}}, \bibinfo {author}
  {\bibfnamefont {Y.~L.}\ \bibnamefont {Chen}}, \bibinfo {author}
  {\bibfnamefont {X.~L.}\ \bibnamefont {Qi}}, \bibinfo {author} {\bibfnamefont
  {Z.~X.}\ \bibnamefont {Shen}}, \bibinfo {author} {\bibfnamefont {S.~C.}\
  \bibnamefont {Zhang}}, \ and\ \bibinfo {author} {\bibfnamefont {L.~W.}\
  \bibnamefont {Molenkamp}},\ }\href {\doibase 10.1103/PhysRevLett.106.126803}
  {\bibfield  {journal} {\bibinfo  {journal} {Phys. Rev. Lett.}\ }\textbf
  {\bibinfo {volume} {106}},\ \bibinfo {pages} {126803} (\bibinfo {year}
  {2011})}\BibitemShut {NoStop}%
\bibitem [{\citenamefont {{Kitaev}}(2009)}]{Kitaev2009}%
  \BibitemOpen
  \bibfield  {author} {\bibinfo {author} {\bibfnamefont {A.}~\bibnamefont
  {{Kitaev}}},\ }in\ \href {\doibase 10.1063/1.3149495} {\emph {\bibinfo
  {booktitle} {American Institute of Physics Conference Series}}},\ \bibinfo
  {series} {American Institute of Physics Conference Series}, Vol.\ \bibinfo
  {volume} {1134},\ \bibinfo {editor} {edited by\ \bibinfo {editor}
  {\bibfnamefont {V.}~\bibnamefont {{Lebedev}}}\ and\ \bibinfo {editor}
  {\bibfnamefont {M.}~\bibnamefont {{Feigel'Man}}}}\ (\bibinfo {address}
  {Chernogolokova},\ \bibinfo {year} {2009})\ pp.\ \bibinfo {pages} {22--30},\
  \Eprint {http://arxiv.org/abs/0901.2686} {arXiv:0901.2686
  [cond-mat.mes-hall]} \BibitemShut {NoStop}%
\bibitem [{\citenamefont {{Ryu}}\ \emph {et~al.}(2010)\citenamefont {{Ryu}},
  \citenamefont {{Schnyder}}, \citenamefont {{Furusaki}},\ and\ \citenamefont
  {{Ludwig}}}]{Ryu2010}%
  \BibitemOpen
  \bibfield  {author} {\bibinfo {author} {\bibfnamefont {S.}~\bibnamefont
  {{Ryu}}}, \bibinfo {author} {\bibfnamefont {A.~P.}\ \bibnamefont
  {{Schnyder}}}, \bibinfo {author} {\bibfnamefont {A.}~\bibnamefont
  {{Furusaki}}}, \ and\ \bibinfo {author} {\bibfnamefont {A.~W.~W.}\
  \bibnamefont {{Ludwig}}},\ }\href {\doibase 10.1088/1367-2630/12/6/065010}
  {\bibfield  {journal} {\bibinfo  {journal} {New J. Phys.}\ }\textbf {\bibinfo
  {volume} {12}},\ \bibinfo {eid} {065010} (\bibinfo {year}
  {2010})}\BibitemShut {NoStop}%
\bibitem [{\citenamefont {Hill}(1994)}]{Hill}%
  \BibitemOpen
  \bibfield  {author} {\bibinfo {author} {\bibfnamefont {T.~L.}\ \bibnamefont
  {Hill}},\ }\href@noop {} {\emph {\bibinfo {title} {The thermodynamics of
  small systems}}},\ \bibinfo {edition} {2nd}\ ed.\ (\bibinfo  {publisher}
  {Dover Publishing},\ \bibinfo {address} {New York},\ \bibinfo {year}
  {1994})\BibitemShut {NoStop}%
\bibitem [{Note1()}]{Note1}%
  \BibitemOpen
  \bibinfo {note} {One can also consider the dependence of the thermodynamics
  on $V$ if it is not a fluctuating parameter, but then one would formally be
  comparing different thermodynamic systems, which is experimentally more
  feasable, but theoretically inelegant.}\BibitemShut {Stop}%
\bibitem [{\citenamefont {Chamberlin}(1999)}]{Chamberlin1999}%
  \BibitemOpen
  \bibfield  {author} {\bibinfo {author} {\bibfnamefont {R.~V.}\ \bibnamefont
  {Chamberlin}},\ }\href {\doibase 10.1103/PhysRevLett.82.2520} {\bibfield
  {journal} {\bibinfo  {journal} {Phys. Rev. Lett.}\ }\textbf {\bibinfo
  {volume} {82}},\ \bibinfo {pages} {2520} (\bibinfo {year}
  {1999})}\BibitemShut {NoStop}%
\bibitem [{\citenamefont {Chamberlin}(2000)}]{Chamberlin2000}%
  \BibitemOpen
  \bibfield  {author} {\bibinfo {author} {\bibfnamefont {R.}~\bibnamefont
  {Chamberlin}},\ }\href {http://dx.doi.org/10.1038/35042534} {\bibfield
  {journal} {\bibinfo  {journal} {Nature}\ }\textbf {\bibinfo {volume} {408}},\
  \bibinfo {pages} {337} (\bibinfo {year} {2000})}\BibitemShut {NoStop}%
\bibitem [{\citenamefont {{Chamberlin}}(2014)}]{Chamberlin2014}%
  \BibitemOpen
  \bibfield  {author} {\bibinfo {author} {\bibfnamefont {R.}~\bibnamefont
  {{Chamberlin}}},\ }\href {\doibase 10.3390/e17010052} {\bibfield  {journal}
  {\bibinfo  {journal} {Entropy}\ }\textbf {\bibinfo {volume} {17}},\ \bibinfo
  {pages} {52} (\bibinfo {year} {2014})},\ \Eprint
  {http://arxiv.org/abs/1504.04754} {arXiv:1504.04754 [cond-mat.stat-mech]}
  \BibitemShut {NoStop}%
\bibitem [{\citenamefont {{Latella}}\ \emph {et~al.}(2015)\citenamefont
  {{Latella}}, \citenamefont {{P{\'e}rez-Madrid}}, \citenamefont {{Campa}},
  \citenamefont {{Casetti}},\ and\ \citenamefont {{Ruffo}}}]{Latella2015}%
  \BibitemOpen
  \bibfield  {author} {\bibinfo {author} {\bibfnamefont {I.}~\bibnamefont
  {{Latella}}}, \bibinfo {author} {\bibfnamefont {A.}~\bibnamefont
  {{P{\'e}rez-Madrid}}}, \bibinfo {author} {\bibfnamefont {A.}~\bibnamefont
  {{Campa}}}, \bibinfo {author} {\bibfnamefont {L.}~\bibnamefont {{Casetti}}},
  \ and\ \bibinfo {author} {\bibfnamefont {S.}~\bibnamefont {{Ruffo}}},\
  }\href@noop {} {\bibfield  {journal} {\bibinfo  {journal} {Phys. Rev. Lett.}\
  }\textbf {\bibinfo {volume} {114}},\ \bibinfo {pages} {230601} (\bibinfo
  {year} {2015})}\BibitemShut {NoStop}%
\bibitem [{\citenamefont {Gibbs}(1878)}]{Gibbs1878}%
  \BibitemOpen
  \bibfield  {author} {\bibinfo {author} {\bibfnamefont {J.}~\bibnamefont
  {Gibbs}},\ }\href@noop {} {\bibfield  {journal} {\bibinfo  {journal} {Transactions
  of the Connecticut Academy of Arts and Sciences}\ }\textbf {\bibinfo {volume} {3}},\
  \bibinfo {pages} {198} (\bibinfo {year} {1878})}\BibitemShut {NoStop}%
\bibitem [{\citenamefont {Di~Francesco}\ \emph {et~al.}(1997)\citenamefont
  {Di~Francesco}, \citenamefont {Mathieu},\ and\ \citenamefont
  {Senechal}}]{DiFrancesco}%
  \BibitemOpen
  \bibfield  {author} {\bibinfo {author} {\bibfnamefont {P.}~\bibnamefont
  {Di~Francesco}}, \bibinfo {author} {\bibfnamefont {P.}~\bibnamefont
  {Mathieu}}, \ and\ \bibinfo {author} {\bibfnamefont {D.}~\bibnamefont
  {Senechal}},\ }\href@noop {} {\emph {\bibinfo {title} {Conformal Field
  Theory}}}\ (\bibinfo  {publisher} {Springer},\ \bibinfo {address} {New
  York},\ \bibinfo {year} {1997})\BibitemShut {NoStop}%
\bibitem [{\citenamefont {Hart}\ \emph {et~al.}(2014)\citenamefont {Hart},
  \citenamefont {Ren}, \citenamefont {Wagner}, \citenamefont {Leubner},
  \citenamefont {M\"uhlbauer}, \citenamefont {Br\"une}, \citenamefont
  {Buhmann}, \citenamefont {Molenkamp},\ and\ \citenamefont
  {Yacoby}}]{Hart2014}%
  \BibitemOpen
  \bibfield  {author} {\bibinfo {author} {\bibfnamefont {S.}~\bibnamefont
  {Hart}}, \bibinfo {author} {\bibfnamefont {H.}~\bibnamefont {Ren}}, \bibinfo
  {author} {\bibfnamefont {T.}~\bibnamefont {Wagner}}, \bibinfo {author}
  {\bibfnamefont {P.}~\bibnamefont {Leubner}}, \bibinfo {author} {\bibfnamefont
  {M.}~\bibnamefont {M\"uhlbauer}}, \bibinfo {author} {\bibfnamefont
  {C.}~\bibnamefont {Br\"une}}, \bibinfo {author} {\bibfnamefont
  {H.}~\bibnamefont {Buhmann}}, \bibinfo {author} {\bibfnamefont
  {L.}~\bibnamefont {Molenkamp}}, \ and\ \bibinfo {author} {\bibfnamefont
  {A.}~\bibnamefont {Yacoby}},\ }\href {http://dx.doi.org/10.1038/nphys3036}
  {\bibfield  {journal} {\bibinfo  {journal} {Nat. Phys.}\ }\textbf {\bibinfo
  {volume} {10}},\ \bibinfo {pages} {638} (\bibinfo {year} {2014})}\BibitemShut
  {NoStop}%
\bibitem [{\citenamefont {Ezawa}\ \emph {et~al.}(2013)\citenamefont {Ezawa},
  \citenamefont {Tanaka},\ and\ \citenamefont {Nagaosa}}]{EzawaGap2013}%
  \BibitemOpen
  \bibfield  {author} {\bibinfo {author} {\bibfnamefont {M.}~\bibnamefont
  {Ezawa}}, \bibinfo {author} {\bibfnamefont {Y.}~\bibnamefont {Tanaka}}, \
  and\ \bibinfo {author} {\bibfnamefont {N.}~\bibnamefont {Nagaosa}},\ }\href
  {http://dx.doi.org/10.1038/srep02790} {\bibfield  {journal} {\bibinfo
  {journal} {Sc. Rep.}\ }\textbf {\bibinfo {volume} {3}},\ \bibinfo {pages}
  {2790} (\bibinfo {year} {2013})}\BibitemShut {NoStop}%
\end{thebibliography}
\end{document}